# Designing a Framework for Smart IoT Adaptations

Asmaa Achtaich [*,1,3], Nissrine Souissi [1,2], Raul Mazo [3,4], Camille Salinesi [3], Ounsa Roudies [1]

1 - Univ. Mohammed V- Rabat, EMI, SIWEB Team - Rabat, Morocco.
2 - ENSMR, Département Informatique - Rabat, Morocco
3 - Université Paris 1 Panthéon-Sorbonne, CRI - Paris, France
4 - Universidad EAFIT - Grupo GIDITIC- Medellin, Colombia
asmaaachtaich@research.emi.ac.ma, roudies@emi.ac.ma, souissi@enim.ac.ma, {raul.mazo,camille.salinesi}@univ-paris1.fr

**Abstract.** The Internet of Things (IoT) is the science of connecting multiple devices that coordinate to provide the service in question. IoT environments are complex, dynamic, rapidly changing and resource constrained. Therefore, proactively adapting devices to align with context fluctuations becomes a concern. To propose suitable configurations, it should be possible to sense information from devices, analyze the data and reconfigure them accordingly. Applied in the service of the environment, a fleet of devices can monitor environment indicators and control it in order to propose best fit solutions or prevent risks like over consumption of resources (e.g., water and energy). This paper describes our methodology in designing a framework for the monitoring and multi-instantiation of fleets of connected objects. First by identifying the particularities of the fleet, then by specifying connected object as a Dynamic Software Product Line (DSPL), capable of readjusting while running.

## 1 Introduction

The Internet of things is a global infrastructure that enables advanced services by interconnecting physical and virtual things like smartphones, sensors, computers, machines, vehicles, buildings, roads, cities or countries, and even people and animals [1]. These services vary from basic context information like location or weather, to much more complex setups. Smart environments are primary applications of the IoT, mainly concerned with issues related to pollution, limited resources, energy optimization, and fault tolerance.

Connected objects can monitor environment indicators like temperature, air and water quality, energy consumption, or radiation. This helps collect information about the surrounding, and prepare solutions to eradicate several phenomenon, or prevent some of the risks. In this context, our work consists of a platform that monitors a fleet of device to preform intelligent and dynamic change for an optimal configuration. When a fleet is implemented, it bears a configuration (FConfig) that is characterized by the set of corresponding devices along with their respective configuration (DConfig).

However, the IoT system is complex, rapidly changing, highly variable, heterogeneous, prone to risks and failure, and extremely dynamic. This implies that in the face of change, the system should have the ability to adapt itself in order to continue offering the needed performance. Dynamic proactive adaptation in particular is required to provide adjustments at runtime [2]. Furthermore, and thanks to IoT devices which are growing exponentially in number and performance, it is much more conceivable to collect real time context data, and react accordingly. Additionally, a Device Management (DM) platform monitors every device in the fleet. It can inspect specific information about the services provided by the device (coffee readiness, light status, expired merchandize, speed of car, motor condition, …), it can collect information about the context of the fleet (temperature, light, location, …) and it can report on the characteristics of the devices themselves (battery life, memory, software version, etc.). In addition to that, and poster to processing the collected data, it is responsible for controlling the fleet in order to adjust its behavior.

In this sense, the paper describes our process in designing a framework for the smart monitoring and reconfiguration of a fleet of connected devices. The paper starts by presenting a motivational example–a smart irrigation fleet, which will be depicted all along the development of our framework. Our process will then be elaborated. The first step identifies the requirements for the management of fleets of connected objects. The second step discusses the particularities of IoT devices and their surroundings. Three representative dimensions are conceived; the system, the context, and the environment. The third step studies the self-adaptation approaches, and selects the Dynamic Software Product Lines (DSPL) paradigm as the mechanism that fits best our set of requirements. The fourth and final step introduces an architecture skeleton; it considers the outcome of the previous stages; the three dimensions on the one hand, and the engineering processes involved in DSPL on the other hand.

The paper is structured as follows: Section 2 presents a motivational example. Section 3 describes our methodology by presenting the requirements needed from the DM platform, describing the characteristics of IoT environments and overviewing the mechanisms for self-adaptation. Section 4 presents the DSPL based framework. And finally, section 5 presents the related works before concluding.

## 2   Motivational examples

In this section, we intend to illustrate the need for proactive self-adaptation of fleets of connected objects. We consider the following irrigation system example: Dust and air humidity sensors, temperature sensors, water sprinklers, water taps, and a smartphone compose a fleet of devices, installed in an agriculture field. Sensors collect data about the dust and air humidity, and about the temperature. When humidity is low, the tap or sprinkler provides dust with the needed water. When the temperature is too high or too low, alerts are sent to the smartphone. The fleet does not take into consideration the specific knowledge related to the domain of agriculture. For instance, instead of watering the plants a days before a rainy day, the fleet could consider the weather forecast to readjust its configuration, and wait for the rain instead of unnecessarily using the water supplies. In this scenario, the proactive adaptation would be possible by implementing a Device Management (DM) Platform that monitors devices and their surroundings, processes the data, and reconfigures the fleet by reconfiguring associated devices.

## 3 Methodology

As we intend to design a framework that manages run-time variability in a fleet of connected objects, the following section outlines our methodology.

### 3.1 Main Requirements Elicitation

In order to insure the proper management of the fleet, the DM is required to provide the necessary mechanisms to monitor IoT devices, to propose best-fit adaptions, to manage different levels of variability and to support a large number of connected devices. Our system's requirements can be identified as follow.

*Smart proactive self-adaptation:* the platform should provide the necessary mechanisms to analyses collected data and adapt the system in problematic situations. In a resources constrained environment like ours, every planned adaptation should be subject to validation to insure its necessity.

*Uncertainty management*: It is not always possible to predict the events that will trigger a reconfiguration. Thus, the platform is required to evaluate the qualities the system offers in comparison with the ones requested by users.

*Variability management*: in a fleet of connected devices, variability can be captured at different levels. The platform should be able to manage this separately throughout the system's lifecycle.

*Physical abstraction:* the platform should support communication with heterogeneous devices and various technologies in order to monitor and actuate. This requirement will not be discussed in this paper. Only preliminary concepts will be introduced.

### 3.2 Identifying Dimensions for IoT Systems

In IoT applications, it is important to take into consideration the mutual dependency between objects and their surroundings -context and environment; change in the surrounding has repercussions on the proper functioning of devices. Similarly, the reconfiguration of the fleet changes the state and behavior of the surrounding. We observed that relevant information comes from three main elements, that we call dimensions. The **system** is the fleet. It is represented by the embedded devices and their configurations. It is managed in a way that its outcome allows the achievement of goals specified by the domain expert. The **context** is everything that surrounds the systems, and has an impact on it. Context is represented by measurements captured by devices that surround the system. Context data can also originate from the user, and it can be time or space bound. Finally, the **environment** illustrates knowledge related to a domain. It holds universal information that might not have a direct impact on the system at a time being. However, it could be significant in other dispositions.

It is important to note that these dimensions are dynamic. Devices that form the system at a particular configuration might not be the same involved in another instance of the same fleet. They could become part of the context. Similarly, information that had an impact on the system in a configuration, might become irrelevant in another, and be part of the environment instead. This confirms the need for variability management. One configuration could correspond to fleet is installed in a covered field during the summer. This installation protects the plants from the burning sun and harmful UV, and helps control the temperature inside the covers. For this installation, the **system** is the water sprinkler, the water tap, and the smartphone. The **context** is the inside temperature, and the dust humidity. And finally, the **environment** is the outside temperature, the weather forecast, the national irrigation laws and the agriculture best practices. During the spring, the field is uncovered. The configuration then switches, the fleet is now installed in an open space. The **system** is still the water sprinkler, the water tap, and the smartphone. The **context** on the other hand now includes the brightness, the air temperature, the dust and air humidity, and the weather forecast. The **environment** contains national irrigation laws and the agriculture best practices. In accordance with these dimensions and with the requirements presented above, a DM platform is required to adjust the fleet to answer the user's needs. The next session discusses self-adaptation mechanisms and selects the best fit for our application.

### 3.3 Selecting a Self-Adaptation Mechanism

A Self-Adaptive Software (SAS) is a system that can automatically modify itself in the face of a changing context, to best answer a set of requirements. The Self-adaption capacity can be provided by programming languages in the form of exceptions, parameters or conditions. However, adaptation through these mechanisms is application specific, error prone and poorly scalable. In contrast to these mechanisms, numerous external approaches contribute to the development of runtime adaptation of software. The following will present an overview of the most notable –but not all- approaches for designing self-adaptive systems.

**Overview of self-adaptive approaches.** Different approaches for SASs can be found in the literature. Reviews and surveys in the matter are available in [3][4]. This section enumerates the most notorious ones, and the design technics they fall into. *Architecture-based* self-adaptive techniques formulate and process changes in an architectural model [5] that describes the properties of software through a set of bound components and interconnections. The two concepts are strictly separated, which allows their rearrangement and replacement. The Rainbow Framework [6] and the three Layer Architecture [7] are the most acclaimed architecture-based approaches for SASs. *Agent-based* approaches model systems as a collection of autonomous agents which can interact within an environment to realize common goals; they create a Multi-Agent System (MAS). In MASs, agents are systems that sense the environment they are part of, and act on it in order to realize a purpose [8]. *Reflection* is the capability of a system to observe and modify its composition at runtime [9]. This technic is used to inspect the internal behavior of a system by implementing additional components for monitoring purposes. It is also used to adapt behavior or structure of a system by changing or replacing or adding features. Reflective middleware like ReIOS [10] are a prominent way to reason about self-adaptation. *Model-driven engineering* (MDE) shifts the focus to the creation and use of domain models, to automate code generation. Models abstract the application and its context, as well as the relationships between them. With regards to self-adaptive systems, MDE provides means for designing manageable systems along with reconfiguration mechanisms to generate executable applications, supported by runtime models during execution [11]. The MUSIC Framework [12] and the Dynamic Software Product Line (DSPL) [13] are model driven approaches. The latter uses models at runtime to address variability and context changes during system execution.

**The DSPL mechanism.** DSPL uses software product lines principles to build systems that can adapt to context fluctuation, new user requirements and variant QoS states. These principles include software reuse, variability modeling and management, and automatic product derivation.

We consider the DSPL paradigm the most fitting approach to provide autonomic scalable support for a fleet of connected devices, from design to execution [14]. First, DSPLs provide a systematic and non-restrictive way to deal with SASs [15], also they successfully realize the MAPE-K loop [16] as tested by Bencomo et al. in [17]. Besides, on the one hand, monitoring and controlling are the main activities for the fleet management. On the other hand, these same two activities are central tasks in DSPLs, which makes the paradigm a good fit for the self-adaptation of the fleet. Also, with regards to uncertainty, the quality of a product can be measured against user requirements by the mean of Goal-based approaches. Goal models can represent the system requirements at the domain level of (D)SPLs, in the form of variable reusable components. Furthermore, variability is a key challenge in the management of a fleet of connected things; it takes place at different levels. Static variability is concerned with similarities and variations between devices, dynamic variability is dealing with the runtime reconfiguration, and temporal variability, describes the alterations of the three dimensions. Dealing with variability is by far the greatest asset of DSPL, since it adopts essential concepts from SPL [18].

The fleets–an irrigation system installed in different fields—can be considered as a DSPL. Each fleet is a product that shares common characteristics with other fleets, but still answers the specific needs of the customer it serves. For instance, some of the devices installed in Sarah's field are like the ones at Omar's. Still, unlike him, Sarah is also interested in measuring the fertility of the soil, and applying fertilizers when needed. A fleet has the capacity to re-adjust itself when requirements are no longer fulfilled. A New FConfig implies a different set of devices with a different DConfig.

## 4   Designing a Fleet as a DSPL

The first level in the process is the creation of assets. As described in Fig. 2, a meticulous study of the domain in question helps define the qualities the system should satisfy, while specifying the variability and the variation points. The result of a domain study is a fleet line (a). The second level is the creation of the final product. The requirements of each customer are described in formal language. The selection of features is carried out accordingly, and then adjusted to fit the exact needs of the customer. Features are finally derived, linked, tested and deployed in order to instantiate the Product—the fleet (d).

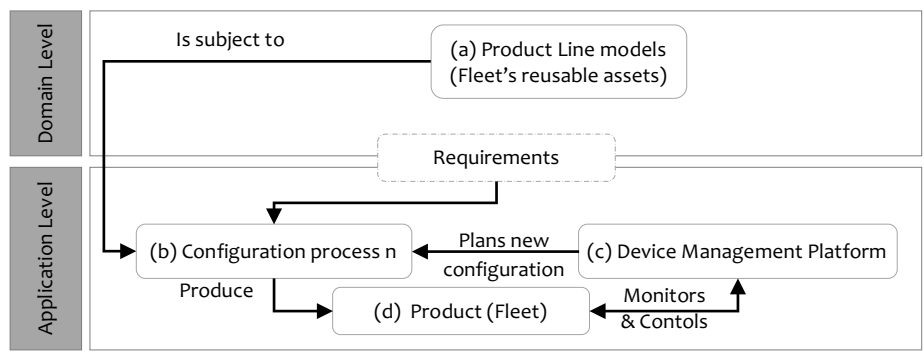

**Fig. 1.** The DSPL Process

DSPLE takes the SPL process one phase further. Each product is thoroughly monitored (c) to determine the structural or behavioral state that dissatisfies requirements. When these are no longer fulfilled, a new configuration is planned (b). This one achieves the optimal satisfaction of primary goals. Features are then re-selected, re-adjusted, re-derived and re-linked (re-tested and re-deployed) to create a new product—a new configuration for the fleet. This process is repeated whenever the system fails to fulfill requirements, in light of contextual change.

From one engineering process to the other, the fleet's three dimensions defined in (3.2) have different designations, as described and illustrated in Fig. 3. At the domain level, each one of the concepts contributes to the creation of assets. With regards to the system (1), a domain expert thoroughly studies the domain in order to determine the functionalities the system should provide and qualities to comply with. In this sense, the system is where domains requirements are extracted, which are then translated to goals, features, components or assets. Context (2) is where the initial requirements are updated to answer the needs that weren't captured by domain experts, but arose after the deployment of the fleet. Environment (3) holds more generic information about domains and devices. It can contribute to the evolution and extensibility of the system by supporting an open Marketplace. This one could supply the system with new components, device specifications, documentation, and other related information.

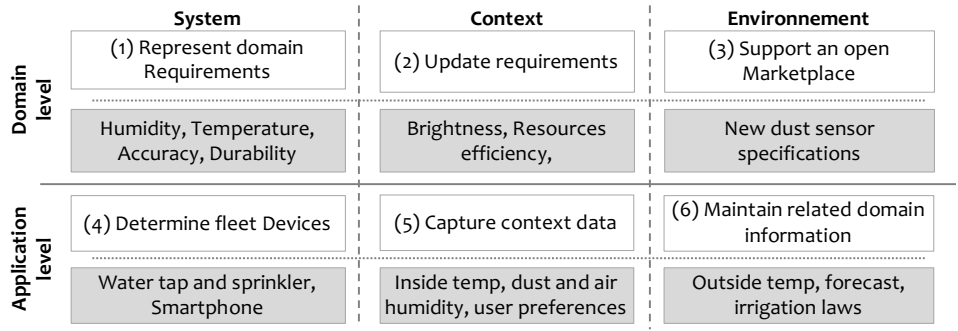

**Fig. 2.** A DSPL three-dimensional Framework

At the **domain level**, each one of the concepts contributes to the creation of assets. With regards to the system (1), a domain expert thoroughly studies the domain in order to determine the functionalities the system should provide and qualities to comply with. In this sense, the system is where domains requirements are extracted, which are then translated to goals, features, components or assets. Context (2) is where the initial requirements are updated to answer the needs that weren't captured by domain experts, but arose after the deployment of the fleet. Environment (3) holds more generic information about domains and devices. It can contribute to the evolution and extensibility of the system by supporting an open Marketplace. This one could supply the system with new components, device specifications, documentation, and other related information. At the **application level**, the monitoring and controlling aspects take place. In relation to the system (4), for each product, devices are monitored in order to determine situations when reconfiguration is required. Sensed or calculated information, feedbacks, battery level, computational performance, network and data accessibility, and other characteristics are relevant. Context (5) on the other hand deals with stakeholders that surround the system, and have an impact on it. Devices that are not part of the system, but contribute to its activity are part of the context, user activity and logs also matter, the time and space of the fleet is also responsible of how it is configured. The environment (6), finally, is place to generic information about the surroundings of the system, that might, but still do not have an impact on the fulfillment of requirements. Devices around the fleet can be in this category, laws, rules or conditions constrained by a time or place are too, part of the environment. Monitoring the environment gives the platform proactive qualities, this helps avoid waste of resources in unnecessary adaptations.

## 5   Related Works

To face the growing complexity of IoT environments, several researchers have identified the need for Frameworks and architectures that support the management of fleets of cooperative devices, considering self-adaptation a core requirement. Inox [19] combines IoT and service architectures to provide enhanced application and service deployment capabilities. The architecture enables the service and network infrastructure with self-management capabilities. In [20], the authors propose an architecture, where agents collect data about protocol operations, measurement-based learning assess the optimality of the control parameter and if necessary, adaptation is realized by applying the new policies to agents. The Focale project [21] introduces an architecture for orchestrating the behavior of heterogeneous distributed resources. Data models support the derivation of different models from a core model, and ontologies reason about the change. The ACE model, proposed in the Cascadas Project [22], defines a agent-based architecture that enables service components to dynamically adapt their behavior based on their context. In [23], a cognitive management framework finds the optimal way to deliver an application in different contexts by enabling the reuse of virtual objects.

With the exception of the Focale Project, none of the above frameworks realize proactive adaptation. Furthermore, in the discussed architectures, no mechanism was proposed to validate the need for intelligent adaptation. Finally, variability is not considered a fundamental concern, thus not managed.

Several SPL based architectures can also be found in the literature. In [24], a DSPL based architecture, combined with preference based reasoning, provides the necessary mechanisms for reasoning about change; this allows the realization of decentralized self-managed system. Gaia-PL [25] is an extension of the Gaia platform for the analysis and design of multi-agent systems in active spaces. A requirement specification pattern captures the behavior of a system in dynamic conditions, and reuses the software assets for future similar systems. In [26], the author proposes a multi-view blueprint architecture, a basis for future smart city projects, based on the SoaSPLE [27] framework for run-time variability management of service-oriented software product lines. Finally, authors in [28] propose a SPL based process for the development of connected devices, defined by the means of CVL, to provide reuse mechanisms for the development of a family of agents.

In contrast with the aforementioned (D)SPL based approaches, our framework introduces variability management at different stages of the process, as explained previously, including static (devices), dynamic (configurations) and time-bound (dimensions alterations) variability. None of the proposed SPL based approaches introduce the environment dimension, necessary for a smart proactive adaptation.

## 6 Conclusion and Perspectives

As a result of a successful COP22 [29], held in 2016 in Marrakech, several Paris agreements were put into practice, including new funds to support climate technologies in developing countries. The IoT paradigm supports this claim by enabling services that manage limited resources, insure service durability, maintain the quality of service, etc. This is possible by supplying connected devices with the necessary mechanisms to readjust their behavior in the face of resource shortage, internet interruptions or service unavailability.

Connected objects can monitor environment indicators, and then a DM Platform processes the information about the surrounding, and prepares solutions to best answer the needs of users. Our work consists of designing a framework for the monitoring and control of a fleet of connected devices, which allows preforming intelligent and dynamic changes for optimal configurations. The first step in our process defines the main requirements needed from the DM platform. The second step defines the characteristics of the fleets, its context and its environment, along with their mutual dependencies. The third step selects DSPL among the various self-adaptation mechanisms as a basis for the framework composition. Considering it is capable of managing uncertainty by capturing inconsistency and readjusting the system's configuration. Eventually, the various modules of the framework are depicted.

This paper has investigated the problem regarding IoT fleets adaptation and proposed a framework for developers to build adaptable applications. Future work includes the validation and implementation of the framework using the VariaMos [30] Tool [31], and an agriculture field case study.

## Acknowledgment

This work was supported by the Moroccan « Ministère de l'Enseignement Supérieur, de la Recherche Scientifique et de la Formation des Cadres », by the « French Embassy in Morocco », and by the « Institut Français du Maroc ».

## References

[1]     I. T. Union, "Overview of the Internet of Things," 2012.
[2]     G. H. Alférez, V. Pelechano, R. Mazo, C. Salinesi, and D. Diaz, "Dynamic adaptation of service compositions with variability models," *J. Syst. Softw.*, vol. 91, no. 1, pp. 24–47, 2014.
[3]     M. Salehie and L. Tahvildari, "Self-adaptive software: Landscape and research challenges," *ACM Trans. Auton. Adapt. Syst.*, vol. 4, no. 2, pp. 1–42, 2009.
[4]     C. Krupitzer, F. M. Roth, S. VanSyckel, G. Schiele, and C. Becker, "A survey on engineering approaches for self-adaptive systems," *Pervasive Mob. Comput.*, vol. 17, pp. 184–206, 2015.
[5]     M. K. Denko, L. T. Yang, and Y. Zhang, "Software Architecture-Based Self-Adaptation," *Auton. Comput. Netw.*, pp. 1–458, 2009.
[6]     S. W. Cheng, D. Garlan, and B. Schmerl, "Evaluating the effectiveness of the rainbow self-adaptive system," *Proc. 2009 ICSE Work. Softw. Eng. Adapt. Self-Managing Syst. SEAMS 2009*, pp. 132–141, 2009.
[7]     J. Kramer and J. Magee, "Self-Managed Systems : an Architectural Challenge," in *Future of Software Engineering*, 2005.
[8]     J. Filipe, A. Fred, and B. Sharp, "Toward a Self-Adaptive Multi-Agent System to Control Dynamic Processes," *Commun. Comput. Inf. Sci.*, vol. 129, 2011.
[9]     E. P. S. Baumer, V. Khovanskaya, M. Matthews, L. Reynolds, V. Schwanda Sosik, and G. Gay, "Reviewing reflection: on the use of reflection in interactive system design," *Proc. 2014 Conf. Des. Interact. Syst. - DIS '14*, pp. 93–102, 2014.
[10]    M. Mongiello, G. Boggia, and E. Di Sciascio, "ReIOS: Reflective Architecting in the Internet of Objects," *Proc. 4th Int. Conf. Model. Eng. Softw. Dev.*, no. February, pp. 384–389, 2016.
[11]    M. Szvetits and U. Zdun, "Systematic literature review of the objectives, techniques, kinds, and architectures of models at runtime," *Softw. Syst. Model.*, vol. 15, no. 1, pp. 31–69, 2016.
[12]    R. Rouvoy, P. Barone, Y. Ding, F. Eliassen, S. Hallsteinsen, J. Lorenzo, A. Mamelli, and U.

Scholz, "MUSIC: Middleware support for self-adaptation in ubiquitous and service-oriented environments," *Lect. Notes Comput. Sci. (including Subser. Lect. Notes Artif. Intell. Lect. Notes Bioinformatics)*, vol. 5525 LNCS, pp. 164–182, 2009.

[13] R. Capilla, J. Bosch, P. Trinidad, A. Ruiz-Cortés, and M. Hinchey, "An overview of Dynamic Software Product Line architectures and techniques: Observations from research and industry," *J. Syst. Softw.*, vol. 91, no. 1, pp. 3–23, 2014.

[14] R. Mazo, C. Dumitrescu, C. Salinesi, and D. Diaz, "Recommendation heuristics for improving product line configuration processes," *Recomm. Syst. Softw. Eng.*, pp. 511–537, 2014.

[15] M. Hinchey, S. Park, and K. Schmid, "Building Dynamic Software Product Lines," *Computer (Long. Beach. Calif).*, vol. 45, no. 10, pp. 22–26, 2012.

[16] IBM, "Autonomic Computing White Paper: An Architectural Blueprint for Autonomic Computing," *IBM White Pap.*, no. June, p. 34, 2005.

[17] N. Bencomo, J. Lee, and S. Hallsteinsen, "How dynamic is your Dynamic Software Product Line?," *Work. Dyn. Softw. Prod. Lines*, 2010.

[18] C. Dumitrescu, R. Mazo, C. Salinesi, and A. Dauron, "Bridging the Gap Between Product Lines and Systems Engineering : An experience in Variability Management for Automotive ... Bridging the gap between product lines and systems engineering . An experience in variability management for automotive model based," in *17th International Software Product Line Conference (SPLC)*, 2013, no. August.

[19] S. Clayman and A. Galis, "INOX: A Managed Service Platform for Inter-Connected Smart Objects Stuart," *Proc. Work. Internet Things Serv. Platforms - IoTSP '11*, pp. 1–8, 2011.

[20] A. Athreya, B. DeBruhl, and P. Tague, "Designing for Self-Configuration and Self-Adaptation in the Internet of Things," *Proc. 9th IEEE Int. Conf. Collab. Comput. Networking, Appl. Work.*, pp. 585–592, 2013.

[21] J. Strassner, N. Agoulmine, and E. Lehtihet, "FOCALE: A novel autonomic networking architecture," *Int. Trans. Syst. Sci. Appl. J.*, pp. 64–79, 2007.

[22] L. Baresi, A. Di Ferdinando, A. Manzalini, and F. Zambonelli., "The CASCADAS Framework for Autonomic Communications," *Auton. Commun.*, no. February 2017, pp. 1–374, 2009.

[23] P. Vlacheas, R. Giaffreda, V. Stavroulaki, D. Kelaidonis, V. Foteinos, G. Poulios, P. Demestichas, A. Somov, A. Biswas, and K. Moessner, "Enabling smart cities through a cognitive management framework for the internet of things," *IEEE Commun. Mag.*, vol. 51, no. 6, pp. 102–111, 2013.

[24] I. Ayala, J. M. Horcas, M. Amor, and L. Fuentes, "Using Models at Runtime to Adapt Self-managed Agents for the IoT," *Sensors*, pp. 155–173, 2015.

[25] J. Dehlinger and R. R. Lutz, "Gaia-PL: A Product Line Engineering Approach for Efficiently Designing Multiagent Systems," *ACM Trans. Softw. Eng. Methodol.*, vol. 20, no. 4, pp. 17:1–17:27, 2011.

[26] M. Abu-Matar, "Towards a software defined reference architecture for smart city ecosystems," *2016 IEEE Int. Smart Cities Conf.*, pp. 1–6, 2016.

[27] M. Abu-Matar and H. Gomaa, "An automated framework for variability management of service-oriented software product lines," *Proc. - 2013 IEEE 7th Int. Symp. Serv. Syst. Eng. SOSE 2013*, pp. 260–267, 2013.

[28] I. Ayala, M. Amor, L. Fuentes, and J. Troya, "A Software Product Line Process to Develop Agents for the IoT," *Sensors*, vol. 15, no. 7, pp. 15640–15660, Jul. 2015.

[29] "COP22." [Online]. Available: http://cop22.ma/en/.

[30] R. Mazo, C. Salinesi, D. Diaz, J. C. Muñoz-Fernández, L. Rincón, C. Salinesi, and G. Tamura, "VariaMos: An Extensible Tool for Engineering (Dynamic) Product Lines," in *Proceedings of the 24th International Conference on Advanced Information Systems Engineering (CAiSE Forum'12)*, 2015, no. June, pp. 374–379.

[31] J. C. Muñoz-Fernández, G. Tamura, M. Raúl, and C. Salinesi, "Towards a Requirements Specification Multi- View Framework for Self-Adaptive Systems," *Comput. Conf. (CLEI), 2014 XL Lat. Am.*, vol. 18, no. 2, pp. 1–12, 2014.